\documentclass[pra,amssymb,twocolumn,superscriptaddress,longbibliography]{revtex4-1}

\usepackage{epsfig}
\usepackage{xcolor}
\usepackage{dcolumn}
\usepackage{graphicx}
\usepackage{hyperref}
\usepackage{graphicx, amsmath, amssymb, times}
\usepackage{siunitx}

\DeclareMathAlphabet{\mathpzc}{OT1}{pzc}{m}{it}

\begin{document}

\title{Mixture of two unequally charged superfluids in a magnetic field}

\author{S. Seyyare Aksu} 
\affiliation{ Department of Physics, Mimar Sinan Fine Arts University, Bomonti 34380, Istanbul, Turkey}
\author{A. Levent Suba{\c s}{\i}}%\email{alsubasi@itu.edu.tr}
\affiliation{Department of Physics, Istanbul Technical University, Maslak 34469, Istanbul, Turkey}
\affiliation{Center for Nonlinear Studies, Los Alamos National Laboratory, Los Alamos, New Mexico 87545, USA}
\author{Nader Ghazanfari} 
\email{nghazanfari@msgsu.edu.tr}
\affiliation{ Department of Physics, Mimar Sinan Fine Arts University, Bomonti 34380, Istanbul, Turkey}

\date{\today}

\begin{abstract}
	The artificial magnetic fields engineered for ultra cold gases depend on the internal structure of the neutral atoms. Therefore the components of a mixture composed of two atomic gases can exhibit a different response to an artificial magnetic field. Such a mixture can be interpreted as a mixture of two atomic gases, carrying different synthetic charges. We consider such mixtures of two superfluids with unequal synthetic charges in a ring trap subject to a uniform artificial magnetic field. The charge imbalance in such a mixture changes the distribution of excited particles over angular momentum states compared to that of an equally charged mixture. This microscopic difference exhibits macroscopic consequences, such as the occurrence of an angular momentum transfer between two unequally charged components. Due to the inter-fluid atomic interactions in a ring, the angular momentum transfer can create a counter flowing persistent current in the weakly charged superfluid. Even in the limiting case of a charged and an uncharged superfluid mixture, a persistent current can be induced in the uncharged superfluid, despite the fact that it is not directly coupled to the magnetic field. The stability analysis shows that the induction depends on the interplay between the inter-fluid interaction and the applied magnetic field. We obtain instability boundaries of the system and construct phase diagrams as a function of the inter-fluid interaction and the magnetic field. We investigate these properties employing the Bogoliubov approximation.
\end{abstract}

\maketitle

\section{Introduction}\label{sec:introduction}
Over the last three decades experimental achievements and developments in the field of ultracold atomic Bose and Fermi gases have provided a wide control over different parameters of these systems. The interaction strength can be finely tuned and different trap geometries can be generated in the experiments for ultracold atomic and molecular systems. Another breakthrough achievement in this field, which came almost a decade ago, has also allowed the coupling of these charge neutral systems to artificial magnetic fields. This coupling becomes possible via a light-induced synthetic magnetic field~\cite{Lin2009a, Lin2009b}. 
\par
Various schemes are used to generate synthetic magnetic fields~\cite{Juzeliunas2006, Zhu2006, Spielman2009, Gunter2009}. However, the general idea is based upon the coupling of a laser beam with the internal states of the atom which gives rise to a geometric phase for the center-of-mass motion. This phase corresponds to an effective vector potential, which is varying in space, and hence can mimic the effect of a magnetic field on a charged particle. The engineered artificial magnetic field depends on the internal structure of the neutral atoms. Therefore, the components of a mixture composed of two different atomic species or of similar atoms with two different hyperfine states can exhibit a different response to an applied artificial magnetic field. In other words, such a mixture can be interpreted as a mixture of two atomic gases, each carrying different synthetic charges subject to an external artificial magnetic field~\cite{Unal2016}. With the prospects of artificial gauge fields, the realization of a mixture of two superfluids with different synthetic charges under an artificial magnetic field is therefore an experimental possibility.
\par
Superfluid mixtures have been studied  and realized in different geometries \cite{Ishino2013, Shimodaira2010, Ticknor2013, Ghazanfari2014, Bandyopadhyay2017,Ramanathan2011, Moulder2012, Murray2013, Beattie2013} including toroidal trapping potentials where the fluid can be modeled as a one-dimensional system on a ring geometry. Different properties of this system have been studied in great detail; persistent currents have been created~\cite{Rokhsar1997b, Bargi2010, Baharian2013, Yakimenko2013, Abad2014, White2016, Gupta2005, Ryu2007, Halkyard2010, Ramanathan2011, Moulder2012, Murray2013, Beattie2013, Wright2013a, Wright2013b, Ryu2013, Neely2013, Jendrzejewski2014, Corman2014, Ryu2014, Eckel2014} and even a long-lived persistent flow of atoms has been observed~\cite{Gupta2005, Moulder2012, Murray2013, Beattie2013}. Moreover, rotational properties and different instabilities of these systems have been scrutinized. However, here we consider a mixture of two unequally charged superfluids for these systems. Motivated by~\cite{Unal2016} and considering the recent experimental realizations of a persistent current in a ring trap, we study the properties of the unequally charged superfluid mixtures.
\par
In a mixture of two unequally charged superfluids, the charge imbalance qualitatively changes the distribution of the excited particles. As a result of this change a highly charged superfluid can transfer the effect of Lorentz force to the weakly charged superfluid via atomic interactions. Thus, the gas which is weakly coupled to the magnetic field can be accelerated by the strongly coupled one. Therefore, a transfer of angular momentum happens between two superfluids. The process of the angular momentum transfer becomes most striking when we consider the limiting case of a charged and an uncharged superfluid mixture. Here, a vortex trapped in the ring, which is the state of a persistent current around the ring, can be generated in the uncharged superfluid despite the fact that it is not directly coupled to the magnetic field. In the case of a ring trap the induction of a persistent current corresponds to a non-dissipative drag as has been demonstrated for some superfluids~\cite{Andreev1975, Tanatar1996, Fil2005}.
\par
The above mentioned angular momentum transfer between two superfluids and hence the possibility of inducing a persistent current to an uncharged superfluid from a charged one can be studied under the Bogoliubov approximation. In the Gross-Pitaevskii approach, where only density-density interactions are considered, the phases of the superfluid wave functions do not enter the energy expression. The absence of these phases prevents any sort of interplay between the velocity fields of the superfluids. Therefore, at the Gross-Pitaevskii level it is not possible to observe an angular momentum transfer between two unequally charged superfluids. Consequently, the induction of a vortex from a charged superfluid to an uncharged superfluid under a magnetic field becomes impossible. However, the Bogoliubov approach that takes excited particles into account allows for an angular momentum transfer.
\par
Following the discussion given above, in this paper we consider a mixture of two superfluids with different synthetic charges in a ring trap subject to a common uniform magnetic field. The paper is organized as follows: In Sec.~\ref{sec:A charged superfluid in a magnetic field}, we review the instabilities and the angular momentum properties of a charged superfluid in a magnetic field. Understanding the properties of a single superfluid will be helpful when we study a mixture of two charged superfluids in the next section. Then, in Sec.~\ref{sec:Mixture of two charged superfluids}, we introduce the mixture of two unequally charged superfluids subject to a magnetic field and obtain the excitation spectrum of the system. The energetic and dynamical instabilities of different mixture types are studied in Sec.~\ref{sec:Instability conditions} from the obtained excitation spectrum. In Sec.~\ref{sec:Angular momentum calculations}, we calculate the angular momentum properties of the superfluid mixtures, and discuss the effect of charge imbalance on these properties. Using the instability analysis illustrated in phase diagrams, the conditions to induce a persistent current from a charged superfluid to an uncharged superfluid is presented in Sec.~\ref{sec:Vortex induction}. Finally, we summarize and discuss our results in Sec.~\ref{sec:Conclusion}. 

\section{A charged superfluid in a magnetic field}\label{sec:A charged superfluid in a magnetic field}
In this section we consider a single component superfluid consisting of $N$ particles each having mass $M$ and synthetic charge which is given as $Q$ times the unit charge $e$. The system is trapped in a quasi-one-dimensional torus geometry with circumference $L=2\pi R$ (with $R$ being the radius of the ring) and cross-section area $S=\pi r^2$, where the tube radius $r \ll R$. The \textit{s}-wave scattering length is assumed to be smaller than the trapping dimensions, i.e. $a_s\ll r,R$, so that the low energy interactions can be derived from the three-dimensional contact potential model as
\begin{equation}\label{eq:coupling}
\mathpzc{g}^{1D} = \frac{4\pi\hbar^2a_s}{M}\frac{1}{S},
\end{equation}
and $\mathpzc{g}^{1D}N/L$ gives the mean interaction energy per particle. The system is under a uniform magnetic field (${\bf B}=B_0\hat{e}_z$) along the central axis of the ring, generated by a vector potential ${\cal A}= \frac{B_0\rho}{2}|_{\rho=R}\hat{e}_\theta$. The many-body Hamiltonian describing this system in the plane-wave basis along the circumference of the ring can be written as
\begin{eqnarray}\label{eq:MB-Ham2-SC}\nonumber
	\mathcal{H} = \sum_{k} \frac{\hbar^2}{2M R^2}\left(k-Q\Phi\right)^2 a^{\dagger}_k a_k && \\ 
	+\; \frac{\mathpzc{g}^{1D}}{2L} \sum_{\substack{k,k',q}} a^{\dagger}_{k}a^{\dagger}_{k'}&&a_{k'-q}a_{k+q}.
\end{eqnarray}
Here, $\Phi = \pi R^2 B_0/\Phi_0$ is the number of the magnetic flux quanta passing through the area enclosed by the ring where $\Phi_0=h/e$ is the magnetic flux quantum. Note that because of the periodic boundary condition the momentum states carry a quantized angular momentum of $\hbar k$, where the quantum number $k$ takes on integer values. The Hamiltonian can be non-dimensionalized using the characteristic energy scale $\hbar^2/(2MR^2)$ which corresponds to the energy difference between the ground and the first-excited state of a single particle in the ring. Accordingly, the dimensionless Hamiltonian becomes
\begin{equation}
	H=\sum_k \left(k-Q\Phi\right)^{2} a^{\dagger}_{k}a_{k}+\dfrac{g}{2}\sum_{k,k^{'},q}a^{\dagger}_k a^{\dagger}_{k'}a_{k'-q}a_{k+q},
\end{equation}
where $g=2MR^2\mathpzc{g}^{1D}/\hbar L=4Ra_s/S$. As mentioned in Sec.~\ref{sec:introduction}, the Bogoliubov approximation provides the mechanism to investigate the excitation spectrum of the system as well as its angular momentum properties. According to the Bogoliubov approach a large fraction of the particles is assumed to be condensed in the lowest-energy single-particle state and the rest are distributed over higher-energy quantum states, so that
\begin{eqnarray}\label{eq:Bogoliubov-approx-SC}
N = \overline{N}_0 + N_{ex} = \langle a^\dagger_{k_0}a_{k_0} \rangle + \sum_{k\neq k_0} a^\dagger_k a_k.
\end{eqnarray}
Here $k_0$ denotes the minimum energy single particle state where the condensate resides and $N_{ex} \ll N_0$. This approximation leads us to a quadratic Hamiltonian of the form
\begin{eqnarray}\label{eq:MB-Ham3-SC}\nonumber
H&=&\left(k_0-Q\Phi\right)^2N+\frac{gN^2}{2} + \sum_{q\neq 0}\left(\epsilon_q+U\right) a^{\dagger}_{k_0+q}a_{k_0+q}\\ 
&+& \frac{U}{2}\sum_{q\neq0}\left(a^{\dagger}_{k_0+q}a^{\dagger}_{k_0-q}+a_{k_0+q}a_{k_0-q}\right),
\end{eqnarray}
for a fixed total number of particles $N$ with $U=gN$. Here $\epsilon_q= q^2 + 2q\left(k_0-Q\Phi\right)$ and $q$ denotes the momentum difference from the condensate momentum $k_0$.  The constant term in the Hamiltonian, i.e. $E_0=\left(k_0-Q\Phi\right)^2N+gN^2/2$ gives the mean field or Gross-Pitaevskii ground state energy of the system. From this energy expression one can obtain a critical magnetic field to generate a persistent current around the ring.  For example, as soon as $Q\Phi>1/2$,  a vortex state with angular momentum $k_0=1$ becomes energetically more favorable against $k_0=0$. Note that this critical magnetic field is determined at the Gross-Pitaevskii level, and the existence of the quantum fluctuations brings a correction to this critical value. 
\par
The improved critical magnetic field in a ring geometry is determined by the energetic instability~\cite{Rokhsar1997a, Anoshkin2013}. This instability is associated with the zeros of the excitation spectrum. The above Hamiltonian becomes diagonal following a usual Bogoliubov transformation, $a_{k_0+q}=u_q \alpha_q - v_q \alpha_{-q}^\dagger$ with the normalization condition $\vert u_q \vert^2 -\vert v_q \vert^2=1$ to ensure that the bosonic commutation relations of the new operators (quasi-particle operators) are conserved. Without going in to the details of the diagonalization procedure, we write the diagonalized Hamiltonian in terms of quasi-particle creation and annihilation operators as
\begin{equation}
H = E_0 + \frac{1}{2}\sum_{q\neq 0} \left(\varepsilon_q -\epsilon_q-U\right)+ \sum_{q\neq 0} \varepsilon_q \alpha^\dagger_q\alpha_q
\end{equation}
where the excitation energy is obtained as
\begin{equation}\label{eq:SC-excitation-spectrum}
\varepsilon_q=2q\left(k_0-Q\Phi\right)+\sqrt{q^4+2Uq^2},
\end{equation}
which differs from the excitation energy of an uncharged superfluid only by the additional term associated with the applied magnetic field. Note that only positive excitations preserve the normalization of the Bogoliubov coefficients $\vert u_q \vert^2 -\vert v_q \vert^2=1$. The occurrence of negative solutions is a direct signature of an energetic instability and indicates that there is a new state with a lower energy. Supposedly, after the energetic instability the superfluid occupies the new state. As seen from Eq.~\eqref{eq:SC-excitation-spectrum} for a charged superfluid in a magnetic field the energetic instability happens when the magnetic field exceeds a certain critical value. After that point, condensate macroscopically occupies a new angular momentum state, which is the $k_0=1$ state. Therefore, the critical magnetic field to excite a persistent current can be readily obtained from equating Eq.~\eqref{eq:SC-excitation-spectrum} to zero for $q=1$, i.e.
\begin{equation}
Q\Phi_c = k_0 + \frac{1}{2}\sqrt{1+2U}.
\end{equation}
\begin{figure}
\includegraphics[width=8.5cm]{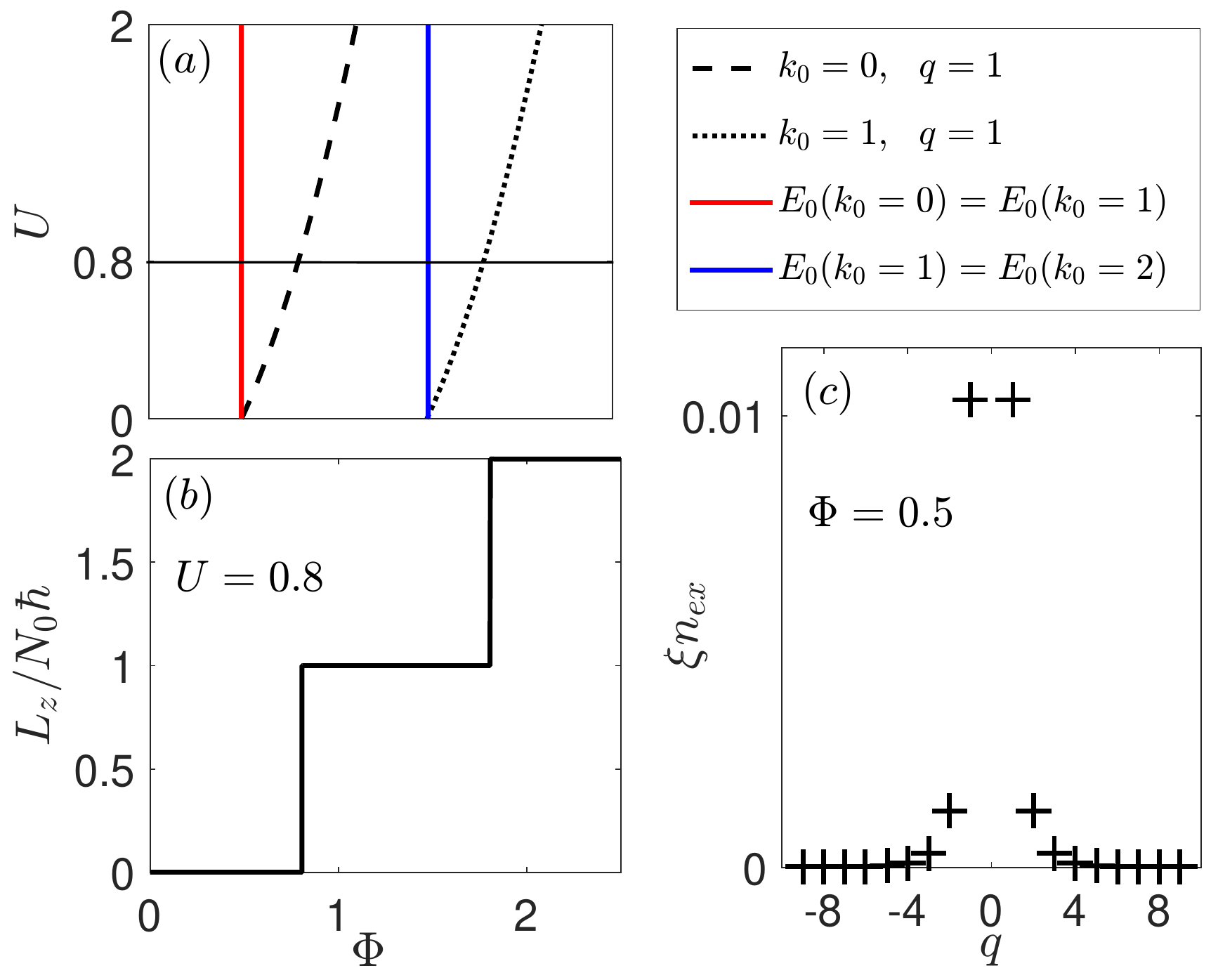}
	\caption{(color online) (a) The dimensionless interaction energy $U$-magnetic flux number $\Phi$ phase diagram of a charged superfluid coupled to an applied magnetic field trapped in a ring-shaped geometry. The vertical lines are associated with the instabilities calculated from the Gross-Pitaevskii energy and the dashed lines show the energetic instabilities obtained from the Bogoliubov excitation spectrum. (b) The angular momentum $L_z/(N_0\hbar)$ is plotted as a function of $\Phi$. At critical $\Phi$ and $U$ values a sudden jump occurs in the angular momentum. (c) The distribution of the excited particles per healing length $\xi=R/\sqrt{U}$ as a function of relative momentum $q$ from the condensate. 
}
\label{fig:SC-phase-diagram}
\end{figure}
\par
Using the excitation spectrum we obtain a phase diagram, as shown in Fig. \ref{fig:SC-phase-diagram}(a), which shows the energetic instability boundaries as a function of the particle interaction strength and the magnetic field values. For very small interactions, the critical magnetic field to excite a persistent current around the ring is close to its value obtained from the Gross-Pitaevskii energy (shown by vertical lines), but the difference becomes greater as the interaction strength increases.
\par
The transition of the condensate from one angular momentum state to another is also signaled by the angular momentum expression which reveals that as soon as the excitation energy becomes zero for a certain $q$ mode, a sudden jump appears in the angular momentum expectation value \cite{Bloch1973}. The angular momentum 
\begin{eqnarray}\label{eq:SC-angular-momentum}\nonumber
\frac{L_z}{\hbar} &=&\sum_{k} k \langle a^\dagger_k a_k\rangle  \\      
&=& k_0 N_0 + \sum_{q\neq 0}(k_0+q) \vert v_q \vert^2(\vert u_q \vert^2 - \vert v_q \vert^2) %\\
	%&=& k_0 N_0 + \sum_{q\ne 0} \frac{k_0+q}{2} \left( \frac{q^2+U}{\sqrt{q^2(q^2+2U)}} -1\right) 
\end{eqnarray}
is affected by the distribution of the non-condensed particles over single particle excited states and the jump happens due to the transition of the condensate to the next angular momentum state. Here 
\begin{equation}
v_q^2 = \frac{1}{2} \left( \frac{q^2+U}{\sqrt{q^2(q^2+2U)}} -1\right). 
\end{equation}
\par
No change in the angular momentum is observed for a superfluid until the condensed particles move to a new vortex state and a persistent current appears inside the superfluid which stays constant until another jump, as seen in Fig. \ref{fig:SC-phase-diagram} (b). Note that the Bogoliubov treatment allows for out of condensate excitations, however the superfluid does not acquire a non-zero contribution to the angular momentum due to the excited particles. Physically, this is because the excitations result from particles scattering out of the condensate with a momentum conserving interaction. Therefore, the distribution of the excited particles over higher angular momentum states as a function of relative momentum from the condensate, as shown in Fig. \ref{fig:SC-phase-diagram} (c), is symmetric and the net contribution to angular momentum from Bogoliubov excitations is always zero. We also note that because of the minimal coupling to the magnetic field, the depletion of the condensate due to interactions is independent of the magnetic field value, until an energetic instability.
\par
From Eq. (\ref{eq:SC-angular-momentum}) we can explicitly confirm that this macroscopic occupation of a new angular momentum state occurs at the energetic instability ($\varepsilon_q=0$). The energetic instability happens when a certain value of magnetic field or interaction energy is exceeded for each angular momentum mode. A larger magnetic field may excite vortices with higher angular momentum quantum numbers~\cite{Anoshkin2013}. In this picture, for a positively charged superfluid subject to a magnetic field in the positive $z$-direction, starting from a no-vortex state with $k_0=0$, no anti-vortex state appears since for negative $q$ values the excitation energy is always positive.

\section{Mixture of two unequally charged superfluids}\label{sec:Mixture of two charged superfluids}

We now consider a mixture of two unequally charged superfluids, represented by operators $a$, and $b$, with corresponding particle masses $M_a$ and $M_b$, and synthetic particle charges $Q_ae$ and $Q_be$, respectively. The mixture is trapped in a ring whose geometry is defined in Sec.~\ref{sec:A charged superfluid in a magnetic field}. The intra-component and inter-component interactions are modeled by the same short-range contact interaction with coupling constants defined as in Eq.~\eqref{eq:coupling}. The system is under a uniform synthetic magnetic field (${\bf B}=B_0\hat{e}_z$) along the central axis of the ring. In the plane wave basis the many-body Hamiltonian describing this mixture can be written as
\begin{eqnarray}\label{eq:MB-Ham}\nonumber
H &=& \sum_{k}\left[ \left( k - Q_a\Phi\right)^2 a^{\dagger}_ka_k + \mu \left( k - Q_b\Phi\right)^2 b^\dagger_k b_k \right] \\ \nonumber
&+&  \frac{1}{2}\sum_{\substack{k,k',q}}\left[g_{a}a^{\dagger}_{k}a^{\dagger}_{k'}a_{k'-q}a_{k+q}+g_{b}b^{\dagger}_{k}b^{\dagger}_{k'}b_{k'-q}b_{k+q}\right] \\ 
&+& g_{ab}\sum_{\substack{k,k',q}}a^{\dagger}_{k}b^{\dagger}_{k'}b_{k'-q}a_{k+q}
\end{eqnarray} 
where $\mu = M_a/M_b$ is the mass ratio between two superfluids and $\Phi = \pi R^2 B_0/\Phi_0$ is the number of magnetic flux quanta passing through the area enclosed by the ring. All the relevant quantities in the Hamiltonian are non-dimensionalized via scaling the Hamiltonian by the characteristic energy $\hbar^2/(2M_aR^2)$ so that $g_a=4R a_s^{(a)} /S$, $g_b=\mu 4Ra_s^{(b)}/S$, and $g_{ab}=(1+\mu)4Ra^{(ab)}_s/S$. It is worth mentioning that by this scaling, the mass ratio information enters into dimensionless coupling constants $g_b$ and $g_{ab}$, and any changes in the mass ratio also implies a change in the coupling constants. We note that the periodic boundary condition imposes integral values for momentum $k$, i.e. $k=\pm 1, \pm 2, \ldots$. 
\par
In order to study the superfluid mixture, we employ the Bogoliubov approximation as in the previous section. According to this approach we assume that
\begin{equation}\label{eq:bogoliubov-approx}
N_a = \overline{N}_a + \sum_{k\neq k_a} a^\dagger_k a_k \quad \mathrm{and} \quad 
N_b = \overline{N}_b + \sum_{k\neq k_b} b^\dagger_k b_k,
\end{equation}
where $k_a$ and $k_b$ are the minimum energy states where condensates $a$ and $b$ reside, respectively. Here, $\overline{N}_a = \langle a^\dagger_{k_a}a_{k_a}\rangle$ and $\overline{N}_b = \langle b^\dagger_{k_b} b_{k_b} \rangle $ are macroscopically large so that the total number of excited particles is very small compared to $\overline{N}_a$ and $\overline{N}_b$, respectively. The Bogoliubov approximation leads us to a quadratic Hamiltonian of the form
\begin{eqnarray}\label{eq:Matrix-Ham}\nonumber
H &=& E_{GP} + \frac{1}{2}  \sum_{q > 0} \left\{\psi^{\dagger}_q \left( {\begin{array}{cc}
    W & Z \\
    Z & W \\
\end{array} } \right) \psi_q - \mathrm{Tr}[W]\right\} \\
	&=& E_{GP}  + \frac{1}{2}  \sum_{q > 0}\left\{ \psi^{\dagger}_q {\cal M} \psi_q - \mathrm{Tr}[W]\right\}
\end{eqnarray} 
where 
\begin{eqnarray}\label{eq:E_GP}
E_{GP} &=& N_a\left(k_a-Q_a\Phi\right)^{2}+\mu N_b\left(k_b-Q_b\Phi\right)^{2}\\ \nonumber
&+& \frac{1}{2}U_a N_a+\frac{1}{2}U_b N_b +U_{ab}\sqrt{N_aN_b}
\end{eqnarray}
is the Gross-Pitaevskii ground state energy with $U_i= g_i N_i$  for $i=a,b$, and $U_{ab}= g_{ab}\sqrt{N_aN_b}$. In Eq.~\eqref{eq:Matrix-Ham} $\psi_q$ is given by the vector operator
\begin{equation}\nonumber
\psi^\dagger_q = \left(a_{k_a+q}^\dagger ~ a_{k_a-q}^\dagger ~ b_{k_b+q}^\dagger~ b_{k_b-q}^\dagger~ a_{k_a+q} ~ a_{k_a-q}~b_{k_b+q}~ b_{k_b-q} \right)
\end{equation} 
and matrices $W$ and $Z$ can be written as
\begin{eqnarray}\nonumber
&&W=
\left( {\begin{array}{cccc}
    \epsilon^a_{q}+U_a & 0 & U_{ab} & 0 \\
    0 & \epsilon^a_{-q}+U_a & 0 & U_{ab} \\
    U_{ab} & 0 & \epsilon^b_{q}+U_b & 0 \\
    0 & U_{ab} & 0 & \epsilon^b_{-q}+U_b \\
 \end{array} } \right),\\
&&Z=
\left( {\begin{array}{*{20}c}
   0 & U_a & 0 & U_{ab} \\
   U_a & 0 & U_{ab} & 0 \\
   0 & U_{ab} & 0 & U_b \\
   U_{ab} & 0 & U_b & 0 \\
 \end{array} } \right)
\end{eqnarray}
where $\epsilon^{i}_q = q^2+ 2q\left(k_i-Q_i\Phi\right)$ for $i=a,b$.
\par
We use the generalized Bogoliubov transformation in order to diagonalize the Hamiltonian and introduce a linear canonical transformation $\psi_q = T \phi_q$, where 
\begin{equation}\nonumber
\phi^\dagger_q = \left(  \alpha_{q}^\dagger ~~\, \alpha_{-q}^\dagger ~~\, \beta_{q}^\dagger ~~\, \beta_{-q}^\dagger ~~\, \alpha_{q} ~~\, \alpha_{-q} ~~\, \beta_{q} ~~\, \beta_{-q} \right)
\end{equation} 
is the vector of quasi-particle creation and annihilation operators. 
The transformation can be explicitly written as 
\begin{align}\label{Bog-transformation}\nonumber
\!\!\!a_{k_a+q} = u_{11}(q)\alpha_{q}\!-\! v_{11}(q)\alpha^{\dagger}_{-q}\!+\! u_{12}(q)\beta_{q}\!-\! v_{12}(q)\beta^{\dagger}_{-q} \\
\!\!\!b^{\dagger}_{k_b+q}= u_{21}(q)\alpha^{\dagger}_{q}\!-\! v_{21}(q)\alpha_{-q}\!+\! u_{22}(q)\beta^{\dagger}_{q}\!-\! v_{22}(q)\beta_{-q}
\end{align}
which brings the Hamiltonian to the following form
\begin{eqnarray}\nonumber
	H &=& E_{GP}+\frac{1}{2}\sum_{q>0}\left\{\phi^{\dagger}_qT^{\dagger}{\cal M} T\phi_q - \mathrm{Tr}[W]\right\}\\ 
	&=& E_{GP} +\frac{1}{2}\sum_{q>0}\left\{\phi^{\dagger}_q\eta T^{-1}\eta {\cal M} T\phi_q - \mathrm{Tr}[W]\right\},
\end{eqnarray}
since for bosons $T^{\dagger}=\eta T^{-1}\eta$. Here 
\begin{equation}
\eta = \left( {\begin{array}{*{20}c} I & 0 \\
0 & -I \\
\end{array}} \right)
\end{equation}
is a diagonal matrix defined by the bosonic commutation relations obeyed by operators, i.e., $\eta_{ij}=\left[\psi_{q,i}, \psi^\dagger_{q,j} \right]$ and $I$ is the $4\times 4$ identity matrix. Note that the new quasi-particle creation and annihilation operators should obey the same commutation relations, which gives 
\begin{eqnarray}\label{norm-relation}\nonumber
&&\vert u_{11}\vert^2  + \vert u_{12}\vert^2 - \vert v_{11}\vert^2 - \vert v_{12}\vert^2 = 1\\
&&\vert u_{21}\vert^2  + \vert u_{22}\vert^2 - \vert v_{21}\vert^2 - \vert v_{22}\vert^2 = 1.
\end{eqnarray}
The transformation leads us to the eigenvalue equation of form $\vert \eta {\cal M} - \lambda I \vert = 0$, or in a more explicit form
\begin{align}\label{eq:eigenvalue-problem}
\left\vert \begin{array}{cccc}
W-\lambda I & Z \\
Z & W + \lambda I \end{array} \right\vert=0.
\end{align}
Consequently, the diagonalized Hamiltonian becomes 
\begin{equation}
H = E_{B} +\sum_{q\neq 0}\left(\varepsilon_q^\alpha\alpha_q^{\dagger}\alpha_q +\varepsilon_q^\beta\beta_q^{\dagger}\beta_q\right).
\end{equation}
where 
\begin{equation}\label{eq:Bogoliubov-ground-state}
 E_{B}=E_{GP} + \sum_{q\neq 0}\left( \varepsilon_q^\alpha + \varepsilon_q^\beta - \epsilon_q^a - \epsilon_q^b - U_a - U_b \right)
\end{equation}
denotes the Bogoliubov ground state energy.
\par
For the general case, there are some methods to diagonalize the system analytically \cite{Sun2010, Anoshkin2013} and the eigenvalue problem in Eq.~\eqref{eq:eigenvalue-problem} can conveniently be solved numerically. The excitation spectra of some mixtures yield simple analytical forms as well which can be used to develop useful insight. For example, the excitation spectrum of an equal mass ($\mu=1$) mixture not subject to a magnetic field is given by 
\begin{equation}\label{eq:ex-spectrum-no-mf}
\varepsilon_q^{\alpha,\beta}\!\!=\!\!\sqrt{\!q^4\! +\! (U_a+U_b)q^2 \pm q^2 \sqrt{4U_{ab}^2 \! +\! (U_a - U_b)^2}}. 
\end{equation} 
This is a well known spectrum for mixtures with an always positive mode and a potentially complex mode for repulsive intra-component interactions. 
\par
In the presence of a magnetic field, we  consider the limit of equal intra-component interactions but different inter-component interactions, i.e. $U_a=U_b=U$, and $U_{ab}\neq U$ with $\mu=1$. We call this case the interaction-balanced case for later reference. The Bogoliubov modes in this case can be written as 
\begin{equation}\label{eq:ex-spectrum-balanced}
\varepsilon_q^{\alpha,\beta}=A_+ + \sqrt{B \pm 2\sqrt{C}}, 
\end{equation}where 
\begin{eqnarray}\label{eq:eigen-parts}
&&\!\!\!\!\!\! A_\pm= q\left[(k_a-Q_a \Phi) \pm (k_b - Q_b \Phi)\right], \\ \nonumber
&&\!\!\!\!\!\! B = q^{4}+2Uq^{2}+ A_-^2, \\ \nonumber
&&\!\!\!\!\!\! C = U_{ab}^2 q^{4} + q^2(2U+q^2) A_-^2.
\end{eqnarray}
The excitations have an always positive mode and a potentially negative or complex mode for repulsive intra-component interactions for which we discuss the relevant parameter regions in the next section. Note that the spectrum reduces to the spectrum of the single component superfluid in a magnetic field given by Eq.~\eqref{eq:SC-excitation-spectrum} by taking $k_a=k_b=k_0$, $Q_a=Q_b=Q$, and $U_{ab}=0$.

\begin{figure}
\includegraphics[width=8.8cm]{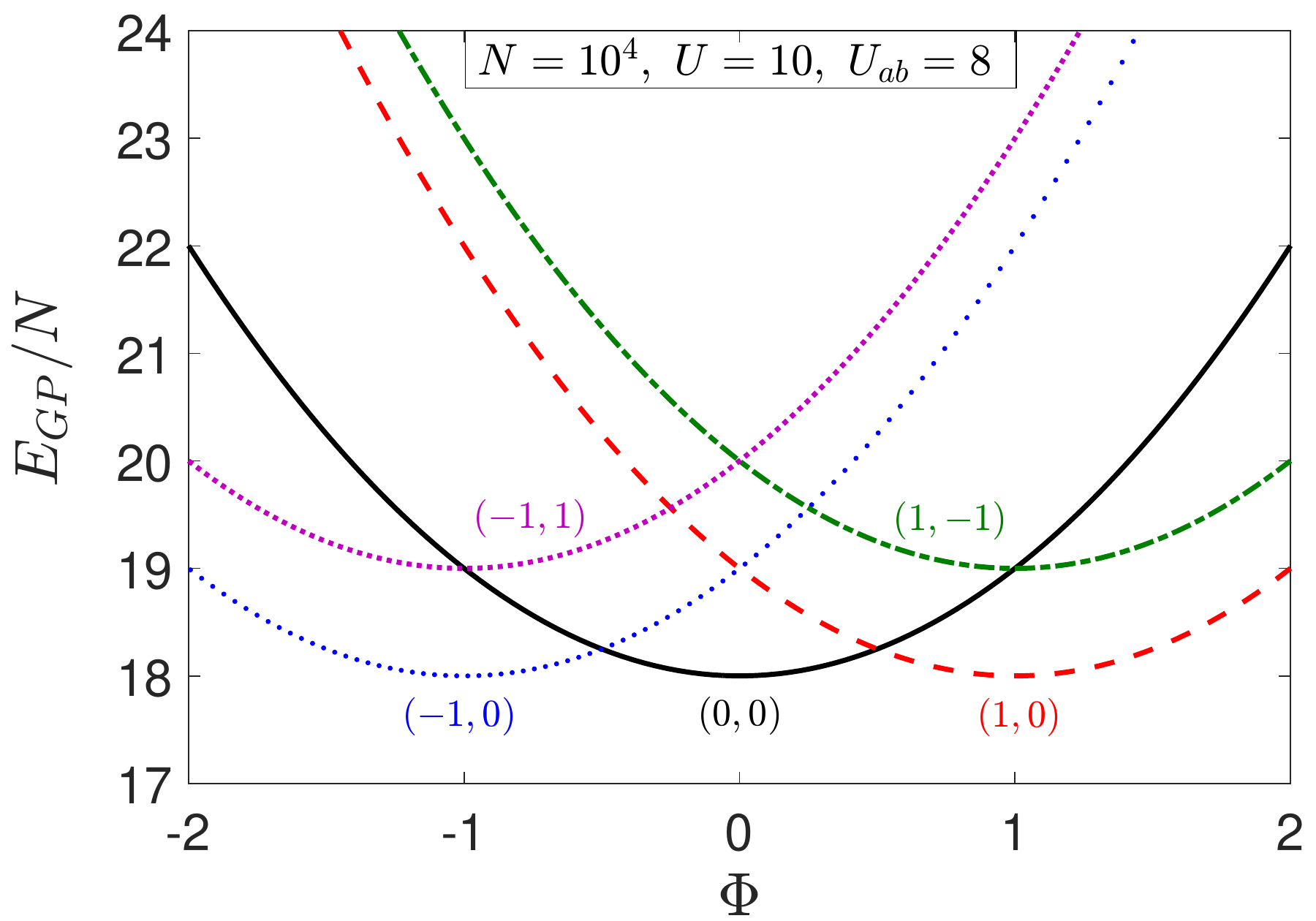}
	\caption{(color online) The Gross-Pitaevskii energies for a charged-uncharged superfluid mixture as a function of magnetic field for different condensates $\left(k_a,k_b\right)$ with equal number of particles $N_a=N_b=N$ and equal intra-component interactions $U_a=U_b=U$. It is seen that the energy of a system with a vortex in one superfluid and a(n) (anti-)vortex in the other is higher than the energy of a system with only a(n) (anti-)vortex in the charged superfluid.
\label{fig:EGP}}
\end{figure}
\par
In Sec. \ref{sec:introduction} we qualitatively discussed that the transfer of angular momentum from a strongly charged superfluid to a weakly charged one is impossible under the Gross-Pitaevskii approximation. This also rules out the possibility of inducing a persistent current from a charged superfluid to the uncharged one. Here, we quantitatively show that Gross-Pitaevskii ground state energies of different superfluids also reveal the same result. From now on, we focus on the interaction-balanced mixtures with charged and uncharged superfluid ($Q_b=0$) components and $N_a=N_b=N$. Figure~\ref{fig:EGP} shows the Gross-Pitaevskii ground state energies of different condensates in $(k_a,k_b)$ angular momentum states as a function of the magnetic field for given particle-particle interactions. It is seen that the energy of a system with a vortex in one superfluid and a(n) (anti-)vortex in the other is higher than the energy of a system with only a(n) (anti-)vortex in the charged superfluid, i.e. ($E_{GP} \left(k_a =\pm 1, k_b=0\right) <  E_{GP}\left(k_a =\pm 1, k_b= \mp 1\right)=E_{GP}\left(k_a =\pm 1, k_b= \pm 1\right)$). This means that observing a persistent current induction from a charged superfluid to an uncharged one is not possible in the context of the Gross-Pitaevskii approximation. Throughout the next section we study the energetic and dynamical instabilities of such a system, and in particular, we determine the critical parameters to obtain a persistent current in the uncharged superfluid using the Bogoliubov modes.

\section{Instabilities of superfluid mixtures} \label{sec:Instability conditions}

There are two kinds of instabilities exhibited by superfluids. The first one is the energetic instability and is observed in all superfluids. As we discussed in Sec.~\ref{sec:A charged superfluid in a magnetic field}, this instability is related to the critical magnetic field value to excite a vortex and is associated with the onset of negative eigenvalues. The second instability, which is called dynamical instability, is characterized by the occurrence of complex eigenvalues in the excitation spectrum~\cite{Pethick2008}. This instability is observed in non-uniform superfluids, such as ultracold atomic condensates, and indicates that such a condensed state is subject to decay. For a single component condensate the dynamical instability is observed with attractive interactions~\cite{Dalfovo1999} which can be identified from the excitation spectrum in Eq.~\eqref{eq:SC-excitation-spectrum} for $U<-1/2$. For two components the dynamical instability signals the spatial separation of the mixture's components~\cite{Pethick2008}. In our instability analysis we concentrate on the two cases for which analytical expressions for the excitation spectrum are given earlier in the previous section. 
\par
First, for the dynamical instability of a mixture of two atomic superfluids with no magnetic field~\cite{Smyrnakis2009, Anoshkin2013} one obtains from Eq.~\eqref{eq:ex-spectrum-no-mf} the condition for which the Bogoliubov excitations become complex valued as
\begin{equation}
\left(U_a+\frac{1}{2}q^2\right)\left(U_b+\frac{1}{2}q^2\right) < U_{ab}^2.
\end{equation}
For the radius $R$ being much greater than the coherence lengths $\xi_i=R/\sqrt{U_i}$ the above relation reduces to $g_a g_b < g_{ab}^2$ for the homogeneous system. For a finite system, the minimum value of inter-component interaction $U_{ab}$ leading to such an instability is greater than that of the homogeneous case. Beyond this critical inter-component interaction, the superfluids constituting the mixture become spatially separated.
\par
Second, we consider the mixture with repulsive intra-component interactions ($U>0$) subject to a magnetic field. Considering Eq.~(\ref{eq:eigen-parts}) it can be easily seen that $C\geq 0$ for all values of interactions, momentum modes, and charges. Moreover, $B\geq 0$ for $U>0$ and we do not consider superfluids with attractive intra-component interactions, i.e $U<0$. Then, one obtains the condition for the onset of the energetic instability as 
\begin{equation}\label{eq:energetic-instability}
A_+ + \sqrt{B - 2\sqrt{C}}<0,
\end{equation}
and the condition for the dynamical instability as
\begin{equation}\label{eq:dynamical-instability}
B < 2\sqrt{C}.
\end{equation}
As seen from the above relations the applied magnetic field splits the onset of the two instabilities. Accordingly, for a given magnetic field value we define two different inter-component interaction values, one associated with energetic instability and the other one associated with the dynamical instability and called $U_{ab}^E$ and $U_{ab}^D$, respectively. These critical inter-component interactions are
\begin{eqnarray}\label{eq:instabilities}\nonumber 
&& U_{ab}^E = \pm \frac{1}{2}\sqrt{\left[A_+^2+A_-^2 -(1+2U) \right]^2+4A_+^2A_-^2},\\
&& U_{ab}^D = \pm \frac{1}{2} \left( 1+2U-A_- \right)
\end{eqnarray}
Alternatively, for fixed interactions, these give implicit equations for two critical values of the magnetic field giving rise to energetic and dynamical instabilities, respectively. These critical values play a very important role in our analysis throughout this paper. At $U_{ab}^E$ a persistent current appears around the ring, and at $U_{ab}^D$ the superfluids become spatially separated. We analyze these two states, i.e., the vortex and the phase separated states, in the following sections.

\begin{figure}
\includegraphics[width=8.5cm]{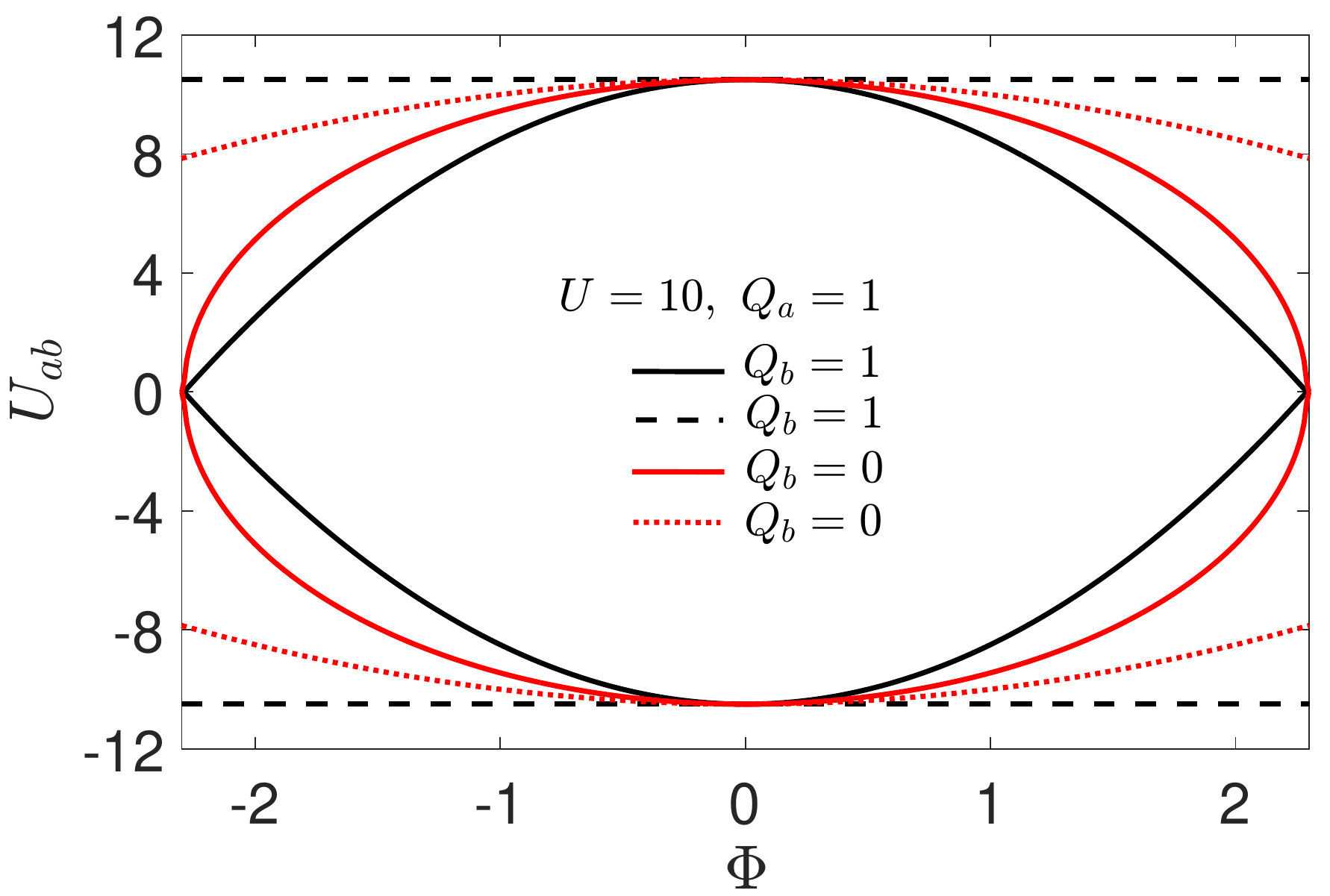}
	\caption{(color online) Phase diagram showing the energetic and dynamical instabilities associated with equally and unequally charged mixtures in the inter-component interaction $U_{ab}$ - magnetic flux number $\Phi$ plane. The critical values $U_{ab}^E$ (solid lines) and $U_{ab}^D$ (dashed lines) are obtained from the onset of zero and complex eigenvalues in the excitation spectrum, respectively.
\label{fig:instabilities}
}
\end{figure}
\par
The energetic instabilities as given in Eq.~\eqref{eq:instabilities} and shown by solid lines in Fig.~\ref{fig:instabilities} are determined by an interplay between the magnetic field and the inter-component interaction. For the equally charged and the charged-uncharged mixtures initially settled at $(k_a,k_b) = (0,0)$ states, it is seen that the phase boundary with a persistent current phase is crossed at smaller magnetic fields for larger inter-component interactions. In other words, with smaller inter-component interaction the energetic instability occurs at a larger magnetic field. Moreover, the charge imbalance enlarges the stability regions of the mixtures in the $\Phi-U_{ab}$. Note that in the presence of a magnetic field the energetic instabilities always happen before the dynamical instabilities shown by the dashed lines in Fig.~\ref{fig:instabilities} indicating a persistent current state before superfluids become spatially separated.
\par
For an equally charged mixture the dynamical instability also given in Eq.~\eqref{eq:instabilities} is independent of the value of magnetic field. This behavior is expected, since in a mixture of two equally charged superfluids, both components feel the same Lorentz force and the transition to phase separation is driven only by increasing the inter-component interactions ($U_{ab}>U$). On the other hand, as a charge imbalance is introduced the dynamical instability becomes magnetic field dependent as shown by the dotted line in Fig.~\ref{fig:instabilities} for $(Q_a=1,Q_b=0)$. The charge imbalance breaks the symmetry between the components under a magnetic field and the phase separation boundary becomes magnetic field dependent.
\par
From Fig. \ref{fig:instabilities} it is also seen that the behavior of the instabilities is symmetric when the magnetic field direction is reversed. A similar symmetry is observed in one-dimensional ring under the exchange of the nature of inter-particle interactions switching from repulsive ($U_{ab}<0$) to attractive ($U_{ab}>0$).

\section{The transfer of angular moementum between two superluids}\label{sec:Angular momentum calculations}
In this section we discuss the effect of the charge imbalance on the angular momentum properties of the mixture. The angular momenta of the components about the axis of the ring are given by
\begin{eqnarray}\label{eq:angular-momentum}\nonumber
&&\!\!\! \frac{L_z^a}{\hbar} \!\!=\!\! \sum_k k \langle a_k^\dagger a_k\rangle = k_a\overline{N}_a + \sum_{q\neq 0} (k_a+q)\langle a_{k_a+q}^\dagger a_{k_a+q}\rangle\\
&&\!\!\! \frac{L_z^b}{\hbar}\!\! =\!\! \sum_k k \langle b_k^\dagger b_k\rangle =  k_b\overline{N}_b + \sum_{q\neq 0} (k_b+q) \langle  b_{k_b+q}^\dagger b_{k_b+q}\rangle.
\end{eqnarray}
and using the Bogoliubov transformation in Eq.~(\ref{Bog-transformation}) can be calculated as
\begin{equation}
\frac{L_z^i}{\hbar} = k_i\overline{N}_i + \sum_{q\neq 0} (k_i+q) \left[ \vert v_{i1}(q)\vert^2 + \vert v_{i2}(q)\vert^2\right],
\end{equation}
for $i= a,b$. The first term above gives the angular momentum of the condensed part and the sum containing the Bogoliubov coefficients $v_{i1}(q)$ and $v_{i2}(q)$ stands for the angular momentum carried by the excited particles. 
\par
In order to bring forth the effects of the charge imbalance, we focus on the interaction-balanced case ($U_{ab}\neq U_a = U_b = U$) and compare the angular momentum properties of the equally charged and the charged-uncharged mixtures. For both types of mixture, we plot the angular momentum distribution of the excited particles relative to the condensates $(k_a,k_b)=(0,0)$ and the total angular momentum of each component in the mixture as a function of the magnetic field in Fig.~\ref{fig:angular-momentum}. 
\par
For an equally charged mixture, the {angular momentum distribution as shown in Fig.~\ref{fig:angular-momentum}(a) is symmetric about $q=0$ for both components, which is similar to the response of a single component superfluid discussed in Sec. \ref{sec:A charged superfluid in a magnetic field} [see Fig.\ref{fig:SC-phase-diagram}(b)]. The collisions giving rise to the depletion of the condensates result from two condensed particles of either condensate scattering into opposite momentum states and these excitations have the same energy when the momenta of the colliding particles are exchanged. Therefore, the net contribution of these excited particles to the total angular momentum is zero and the total angular momentum of each superfluid remains constant and equal to the angular momentum of the condensed part as shown in Fig.~\ref{fig:angular-momentum}(b). 
\par
We note in passing that the symmetric picture discussed above remains qualitatively the same when there is an imbalance between the intra-component interactions or the masses (the mass ratio $\mu$ between two superfluids is also inside the scaled parameters $U_a$ and $U_b$, as mentioned in Sec.~\ref{sec:Mixture of two charged superfluids}), i.e. the distributions remain symmetric about $q=0$ but the depletion in each component may be different.

\begin{figure}
\includegraphics[width=8.8cm]{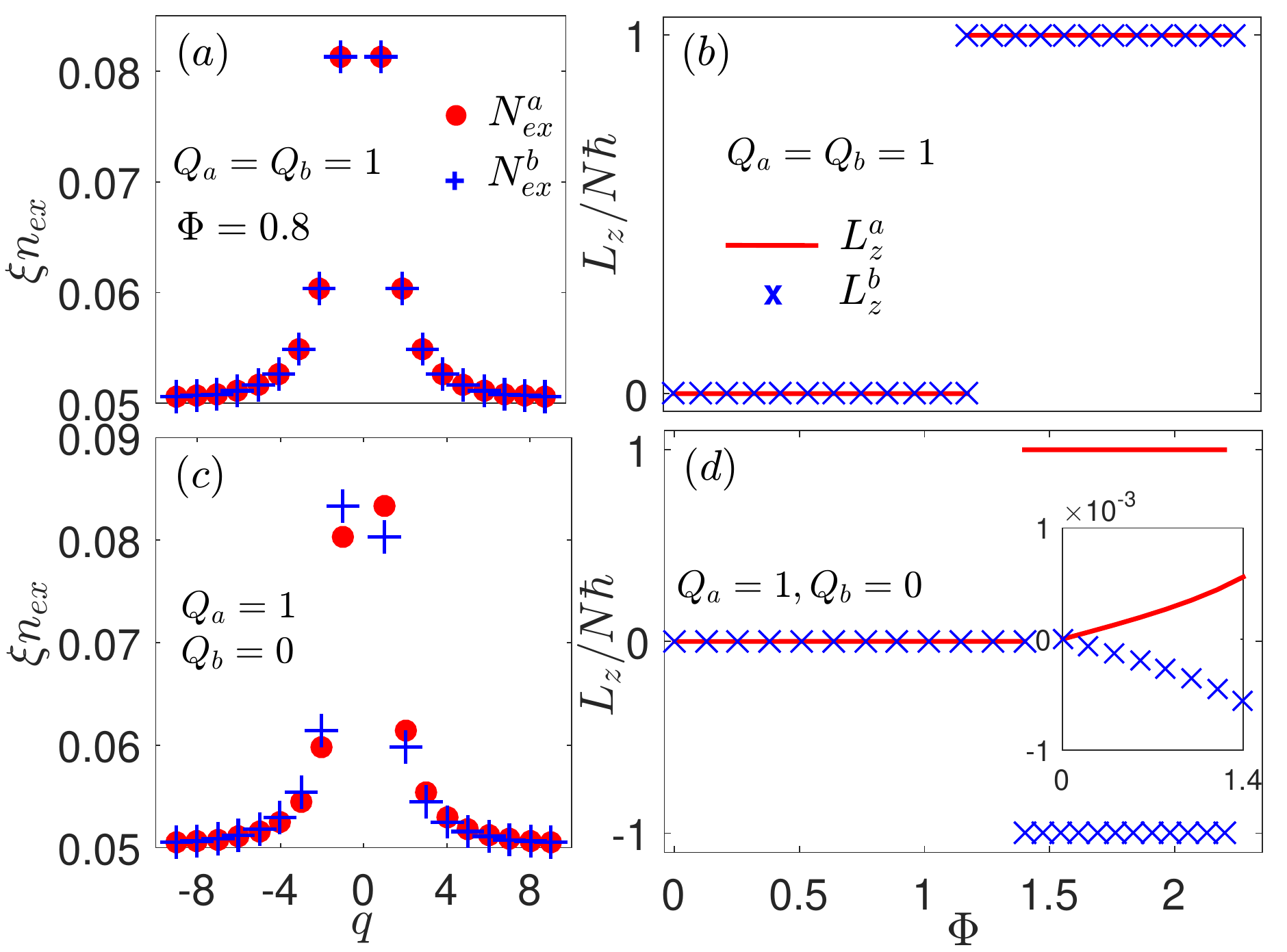}
\caption{
(color online) The angular momentum distribution of excited particles per healing length $\xi$ for charge-balanced and charge-imbalanced mixtures for equal intra-component interactions $U=10$ and inter-component interaction $U_{ab}=8$ are plotted in (a) and (c), respectively. The charge imbalance breaks the symmetry about $q=0$ for $(k_a,k_b)=(0,0)$ mixture. The total angular momentum of the components of the two mixtures for $U=10$ and $U_{ab}=8$ are plotted as a function of the magnetic flux in (b) and (d), respectively. The sudden jump at the critical values indicates the creation of persistent currents. For the charge-imbalanced mixture, the difference in the excitation spectrum leads to finite angular momenta for both components before the critical field and counter-flowing persistent currents after the critical field as shown in the inset in (d).}
\label{fig:angular-momentum}
\end{figure}
\par
For two unequally charged superfluids, we note that the inter-component interaction causes a qualitative change in the distributions of non-condensed particles of both components compared to those of two equally charged superfluids. The charge imbalance breaks the symmetry of excitations. The resulting angular momentum distributions are shown in Fig. \ref{fig:angular-momentum}(c). In this case, the nature of excitations is again such that when a particle from the highly charged superfluid is excited, another one from the weakly charged superfluid gets excited to an angular momentum state with the same magnitude, but opposite direction. However, excited particles from the highly charged superfluid rotate more in the same direction with the field and excited particles from the weakly charged superfluid more in the opposite direction which explains the asymmetry of the momentum distributions about $q=0$ for each superfluid.
\par
As a result of the asymmetry in the momentum distributions, the net relative angular momentum of the excited particles in each superfluid becomes non-zero and proportional to the applied magnetic field. This is shown in the inset in Fig.~\ref{fig:angular-momentum}(d) as a function of increasing magnetic field. Unlike the charge balanced case, the total angular momentum of each superfluid changes when a magnetic field is applied. We can therefore interpret the result as a transfer of angular momentum from the highly charged superfluid to the weakly charged one albeit in the opposite direction (an anti-drag effect). The total angular momentum of the system is conserved as the excess angular momenta of both superfluids cancel each other.
\par
We check the validity of the Bogoliubov approximation 
by calculating the number of excited particles $N_\mathrm{ex}$ to ensure 
that this number stays small with respect to the total number of particles. 
It is convenient to express this number per coherence length $\xi = R /\sqrt{U}$, 
%which involves the same product of the scattering length and the number of particles as it appears in the dimensionless interactions strength $U$, 
and it should be smaller than one~\cite{Pethick2008}.
The density of the excited particles can be written as
\begin{equation}
n_\mathrm{ex}^i=\! \frac{1}{2\pi R}\sum_{q\neq 0} \left[ \vert v_{i1}(q)\vert^2 + \vert v_{i2}(q)\vert^2\right]\!= \frac{1}{2\pi \sqrt{U}}\frac{N^i_\mathrm{ex}}{\xi}.
\end{equation} 
We note that the excitation number does not depend on the magnetic field for equally charged mixtures because of the symmetry between the dispersion relations of the components. For unequally charged mixtures there is a slight dependency as the symmetry between the components is broken. 
For example, for $U=10$ we find $N_\mathrm{ex}\simeq 2$ for both components until the energetic instability which corresponds to $\xi n_\mathrm{ex} \simeq 0.1$. 
Assuming a scattering length of $a_s=\SI{10}{\nano\m}$, $U=10$ requires $N=10^3$ and results in a fraction of $N_\mathrm{ex}/N\simeq\SI{2e-3}{}$. The distribution of excited particles per healing length $\xi$ is shown in Figs.~\ref{fig:angular-momentum}(a), and \ref{fig:angular-momentum}(c). 
For the same scattering length and $U=100$, the corresponding numbers are $N=10^4$, $N_\mathrm{ex}\simeq 12$, $\xi n_\mathrm{ex}\simeq 0.2$ and $N_\mathrm{ex}/N\simeq\SI{1.2e-3}{}$, respectively. 
\par
It is worth mentioning that the behavior of the instability conditions and the angular momentum transfer between two superfluids in a one-dimensional ring remains unchanged when the sign of the inter-component interaction is reversed. The attractive interaction acts similarly to the repulsive one. We believe that considering the finite width of the superfluids will change this picture, which will be the subject of our future studies.

\section{Induction of a persistent current from a charged superluid to an uncharged superfluid}\label{sec:Vortex induction}

From angular momentum calculations of both equally and unequally charged superfluid mixtures presented in Figs. \ref{fig:angular-momentum}(b) and \ref{fig:angular-momentum}(d), we observe that after a critical value of the magnetic flux $\Phi$, a sudden and simultaneous change occurs in the angular momentum of both components. This sudden macroscopic change in the angular momentum, which happens at the energetic instability of the mixture, shows the appearance of persistent currents. Consequently, the initial mean field collapses and the calculations should be done with a new mean field, i.e. this time the superfluids residing in new $k_a$ and $k_b$ values, until the next instability.
\par
The process of exciting persistent currents in both components in a charged-uncharged superfluid mixture is interesting. Here, the angular momentum transfer creates a persistent current in the uncharged superfluid which is not directly coupled to the applied magnetic field. In other words, this is the induction of a vortex from a charged superfluid coupled to a magnetic field to an uncharged superfluid. The uncharged superfluid indirectly becomes coupled to the magnetic field via the interactions between two superfluids. As a result, two counter-flowing currents can appear in the ring. In order to discuss the conditions for the induction of a persistent current in an uncharged superfluid we need to explore the stability diagrams of the mixtures in more detail.
\begin{figure}
\includegraphics[width=8.5cm]{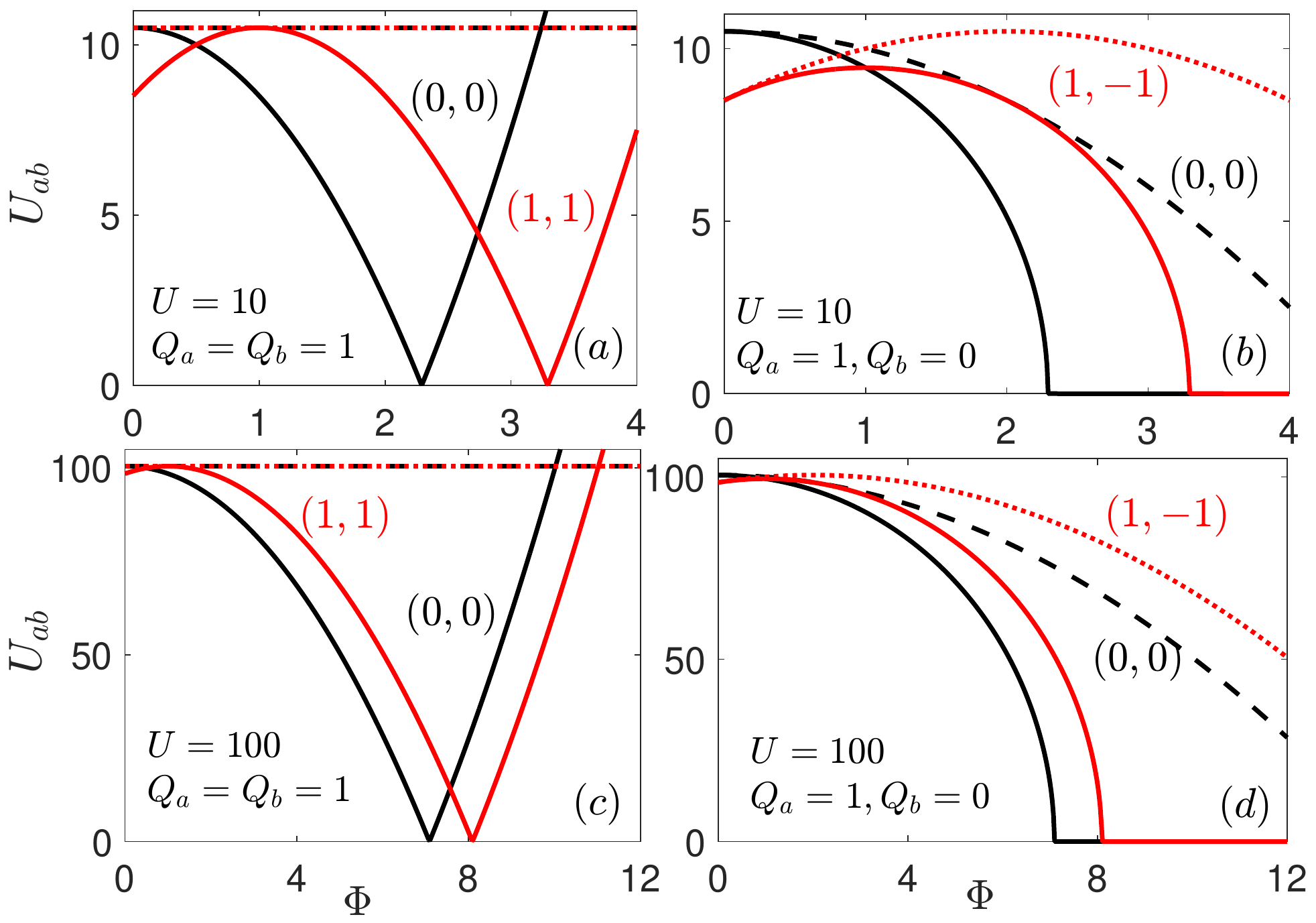}
\caption{(color online) 
	The energetic (solid lines) and dynamical (dashed lines) instability boundaries for different condensates $(k_a,k_b)$ of equally (left column) and unequally (right column) charged superfluid mixtures in terms of the inter-component interactions $U_{ab}$ and the magnetic flux number $\Phi$. For unequally charged mixtures, a stable region for two counter-flowing superfluids exists beyond the energetic instability of the initially at rest superfluids. The increase in the magnitude of the intra-component interaction from $U=10$ (upper panels) to $U=100$ (lower panels) shifts the energetic instability of the mixture to higher magnetic field values while the qualitative behavior remains the same.}
\label{fig:phase-diagrams}
\end{figure}
\par
The phase diagrams in Fig.~\ref{fig:phase-diagrams} show the energetic (solid lines) and dynamical (dashed lines) stability boundaries 
determined by the interplay between the inter-component interaction strength and the magnetic field for charge balanced and charge imbalanced mixtures, respectively. The dynamical instability as we discussed in Sec. \ref{sec:Instability conditions} for a mixture of two equally charged superfluids is independent of the applied magnetic field and appears when a certain value of the inter-component interaction is exceeded. For a mixture of two unequally charged superfluids this instability becomes magnetic field dependent. However, it always appears after the energetic instability.
\par 
For both equally (Fig.~\ref{fig:phase-diagrams}, left column) and unequally (Fig.~\ref{fig:phase-diagrams}, right column) charged mixtures the instability boundaries show that the mixture of two superfluids with a persistent current created in each is locally stable. For a mixture of two equally charged superfluids this stable region corresponds to the state of two persistent currents flowing in the same direction, i.e. two superfluids residing in $(k_a,k_b)=(1,1)$ angular momentum states. However, for a mixture of a charged and an uncharged superfluid, a state of two counter-flowing superfluids, i.e. superfluids occupying the states $(k_a,k_b)=(1,-1)$ is stable. This phase which emerges as a result of inducing a persistent current to the uncharged superfluid from charged superfluid coupled to the magnetic field happens before the dynamical instability of the initial non-rotating mixture and has energetic and dynamical stability. Then, the induction of two persistent currents happens before superfluids are spatially separated.

The parameters used to obtain these results are compatible with the parameters used in recent experiments. The radius of the ring $R$ ranges from 
\SIrange{12}{25}{\micro\meter} and the cross-sectional radius $r$ of the toroidal trapping potential is around \SI{5}{\micro\meter}~\cite{Henderson2009, Ramanathan2011, Beattie2013, Jendrzejewski2014, Ryu2014}. An intra-component interaction strength of $U=10$ with $N=10^3$ particles corresponds to an \textit{s}-wave scattering length of \SI{10}{\nano\meter}, which is reasonable for the experimental setups. Note that the qualitative behavior of the results presented is not very sensitive to the strength of the intra-component interaction (compare the upper and lower panels in Fig.~\ref{fig:phase-diagrams}). The increase in the magnitude of the intra-component interaction only shifts the energetic instability of the mixture to higher magnetic field values. Therefore, a scaled interaction strength of $U=100$ with $N=10^4$ particles corresponds to an \textit{s}-wave scattering length of again \SI{10}{\nano\meter}, however with an instability predicted at a different magnetic field.  

\section{Conclusion}\label{sec:Conclusion}

We considered a mixture of two unequally charged superfluids in a ring trap subject to a uniform magnetic field. The charge imbalance in this mixture causes qualitative changes in the excitation properties of the system compared to those of a two equally charged superfluid mixture. In a mixture of two equally charged interacting superfluids, the distribution of the excited particles over angular momentum states for each component is similar to that of a single component superfluid. Due to the symmetry between the components, the excited particles are distributed symmetrically with respect to the condensate momenta and carry zero net angular momentum. The excitation of persistent currents is determined by the internal dynamics of each component in this mixture. 
\par
For a mixture of two unequally charged superfluids the distribution of the excited particles over angular momentum states changes completely which causes a change in the angular momentum properties of the system. In this case, we calculate a finite angular momentum associated with the contribution of the excited particles to the total angular momentum of each superfluid. These finite angular momenta are equal in magnitude but opposite in direction so that the total angular momentum of the system is conserved. Moreover, a finite angular momentum appears in both superfluids even when one of the superfluids is uncharged, and hence not directly coupled to the magnetic field. In other words, we show that a transfer of angular momentum occurs from a charged superfluid to an uncharged one. This transfer of angular momentum causes counter-flowing persistent currents around the ring. 
\par
The conditions for the persistent current induction are studied through the instabilities of the mixtures. According to the instability analysis the induction of a persistent current from a charged superfluid to an uncharged one is allowed energetically and dynamically. The persistent currents flow in the opposite directions regardless of the attractive or repulsive nature of the inter-component interactions, since for a one-dimensional ring geometry the attractive inter-component interaction acts similarly to the repulsive one. We believe that considering a two-dimensional harmonic trap will change this picture, which will be the subject of our future studies. 

\acknowledgements

The authors thank M.~\"O. Oktel for initial motivation and useful discussions. This work is supported by T\"{U}B\.{I}TAK under Project No. 117F469. A.L.S. is supported by the T\"{U}B\.{I}TAK B\.IDEB 2219 scholarship program and by the Laboratory Directed Research and Development program of Los Alamos National Laboratory under Project No. 20180045DR.

\bibliography{BEC}

\end{document}